\documentclass[aps,prb,amsmath,amssymb,a4paper,reprint,nofootinbib,showkeys]{revtex4-2}

\usepackage{verbatim}
\usepackage{graphicx}
\usepackage[utf8]{inputenc}
\usepackage{hyperref}
\usepackage{epstopdf}
\usepackage{listings}
\usepackage{xcolor}
\usepackage{array}

\newcommand{\cyan}{\color{cyan}}
\newcommand{\be}{\begin{equation}}
\newcommand{\ee}{\end{equation}}
\newcommand{\md}{\mathrm{d}}
\newcommand{\nn}{\nonumber}
\newcommand{\eF}{\epsilon_\mathrm{_F}}
\newcommand{\Fref}[1]{Fig.~\ref{#1}}
\newcommand{\Eqref}[1]{Eq.~(\ref{#1})}

\newcommand{\REM}[1]{}

\begin{document}

\title[Hot spots by \textit{s-d} exchange interaction]
{Hot spots along the Fermi contour of high-$T_c$ cuprates analyzed by 
\textit{s-d} exchange interaction}

\author{Todor~M.~Mishonov}
\email{mishonov@bgphysics.eu}
\affiliation{Georgi Nadjakov Institute of Solid State Physics, Bulgarian
Academy of Sciences, 72 Tzarigradsko Chauss\'ee, BG-1784 Sofia, Bulgaria}

\author{Nedelcho~I.~Zahariev}
\email{zahariev@issp.bas.bg}
\affiliation{Georgi Nadjakov Institute of Solid State Physics, Bulgarian
Academy of Sciences, 72 Tzarigradsko Chauss\'ee, BG-1784 Sofia, Bulgaria}

\author{Hassan Chamati}
\email{chamati@issp.bas.bg}
\affiliation{Georgi Nadjakov Institute of Solid State Physics, Bulgarian
Academy of Sciences, 72 Tzarigradsko Chauss\'ee, BG-1784 Sofia, Bulgaria}

\author{Albert~M.~Varonov}
\email{varonov@issp.bas.bg}
\affiliation{Georgi Nadjakov Institute of Solid State Physics, Bulgarian
Academy of Sciences, 72 Tzarigradsko Chauss\'ee, BG-1784 Sofia, Bulgaria}

\date{1 August 2022, 17:56}

\begin{abstract}
We perform a thorough theoretical study of the electron properties of a generic
CuO$_2$ plane in the framework of Shubin-Kondo-Zener \textit{s-d} exchange interaction
that simultaneously describes the correlation between \textit{T}$\!_c$
and the Cu4\textit{s} energy. To this end,
we apply the Pokrovsky theory \href{http://jetp.ras.ru/cgi-bin/dn/e_013_02_0447.pdf}{[J. Exp. Theor. Phys. 13, 447-450 (1961)]} for anisotropic gap BCS superconductors.
It takes into account the thermodynamic fluctuations of the electric field
in the dielectric direction perpendicular to the conducting layers.
We microscopically derive a multiplicatively separable kernel
able to describe the scattering rate in the momentum space, as well
as the superconducting gap anisotropy within the BCS theory.
These findings may be traced back to the fact that both the Fermi
liquid and the BCS reductions lead to
one and the same reduced Hamiltonian involving a separable
interaction, such that a strong electron scattering corresponds to a strong superconducting gap
and vice versa. Moreover, the superconducting gap and the scattering
rate vanish simultaneously
along the diagonals of the Brillouin zone.
We would like to stress that our theoretical study reproduces the
phenomenological analysis of other authors
aiming at describing Angle Resolved Photoemission
Spectroscopy measurements.
Within standard approximations one and the same 
\textit{s-d} exchange Hamiltonian describes gap anisotropy
of the superconducting phase and the anisotropy of 
scattering rate of charge carriers 
in the normal phase.
\end{abstract}


\maketitle

\section{Introduction}

Since the discovery of high temperature superconductivity, 
the cuprates are one of the most studied materials.
Nevertheless, the theoretical challenge to predict the critical
temperature, say $T_c$,
of certain materials still remains open~\cite{Gozar:16} despite
the significant number of theoretical studies on this topic.\footnote{This work has to be cited  as Ref.~\cite{hotspot2022}}
The purpose of our work
is to give a microscopic explanation of hot/cold spots phenomenology
\cite{Hlubina:95,Ioffe:98}
of the normal phase of the optimally and over doped cuprates
using Shubin-Kondo-Zener \textit{s-d} exchange interaction in the CuO$_2$ 
which allows to explain $T_c$--Cu$4s$ energy correlation~\cite{Pavarini:01}.
Hot spots are the regions with strong scattering and short lifetime,
while cold spots are the regions with the longest lifetime or the weakest scattering.
R\"ohler~\cite{Roehler:00} noted that 
the Cu$4s$-3$d_{x^2-y^2}$ hybridization seems to be the crucial quantum chemical parameter controlling related electronic degree of freedom.
For optimally doped and overdoped cuprates the used LCAO (linear combination of atomic orbitals) approximation for the electron bands~\cite{MishonovPenev:11} agrees with Local Density Approximation (LDA) band 
calculations~\cite{Andersen:95,Andersen:96} and 
Angle Resolved Photoemission Spectroscopy (ARPES) measurements
\cite{Shen:95,Randeria:97}.
The named experiments on optimally doped cuprates
demonstrated that for momenta parallel to $(\pi,\,\pi)$,
the electron spectral function 
exhibits a reasonably well defined quasiparticle peak, suggesting relatively weak scattering.

In this paper,
we provide a microscopic derivation of this phenomenology
devised in Refs. \cite{Hlubina:95,Ioffe:98}.
The best conditions for the applicability of this phenomenology
are satisfied for
optimally and overdoped systems of layered cuprates.
To achieve our task, we use the life-time $\tau_\mathbf{p}$, 
mean free path $l_\mathbf{p}$, and other quantities of the 
kinetics of the normal phase when we describe a normal metal.
This is the case of optimally and overdoped cuprates 
far from Mott~\cite{Mott:90} metal-insulator
transition and negligible, if any, pseudogap.
In short, hot/cold spot phenomenology is applicable roughly speaking
when cuprates exhibit a Fermi liquid behavior.
Then, we can apply a standard set of notions for
normal metals~\cite{Abrikosov}.
There is an emerging consensus that high-$T_c$ superconductivity
is generated by an exchange interaction
and in Section \ref{sd}, we introduce the simplest exchange interaction compatible with the gap anisotropy.
In Section \ref{notions}, we briefly present the theoretical details, notations and notions
of the \textit{s-d} exchange interaction following Ref.~\cite{MishonovPenev:11}.
ARPES experiments measure hot and cold spots regions
accurately as it is shown in Section~\ref{sec:Res}.

\section{The $s$-$d$ LCAO Hamiltonian}\label{sd}
We start with the Linear Combination of Atomic Orbitals (LCAO)
Hamiltonian~\cite[Eqs.~(1.2), (2.9) and 
Figs.~1.1, 2.1]{MishonovPenev:11}
\begin{align}
\hat{H}^\prime = & \sum_{\mathbf{n},\alpha} \left\{ 
\hat D_{\mathbf{n},\alpha}^\dagger \left[ (\epsilon_{d}-\mu)\hat
D_{\mathbf{n},\alpha} \right.\right.  \notag \\
 & \left. -t_{pd}
\left(-\hat X_{\mathbf{n},\alpha}+\hat X_{x-1,y,\alpha}
+\hat Y_{\mathbf{n},\alpha}-\hat Y_{x,y-1,\alpha}\right)\right]\nn\\&
+\hat S_{\mathbf{n},\alpha}^\dagger\left[\left(\epsilon_{s}-\mu\right)\hat
	S_{\mathbf{n},\alpha}\right.\nn\\&
-t_{sp}
	\left.\left(-\hat X_{\mathbf{n},\alpha}+\hat X_{x-1,y,\alpha}
-\hat Y_{\mathbf{n},\alpha}+\hat Y_{x,y-1,\alpha}\right)\right]\nn\\&
+\hat X _{\mathbf{n},\alpha}^\dagger [-t_{sp}(-\hat S_{\mathbf{n},\alpha}+\hat S_{x+1,y,\alpha})\nn\\&
-t_{pp}(\hat Y_{\mathbf{n},\alpha}-\hat Y_{x+1,y,\alpha}
-\hat Y_{x,y-1,\alpha}+\hat Y_{x+1,y-1,\alpha})\nn\\&
-t_{pd}(-\hat D_{\mathbf{n},\alpha}+\hat D_{x+1,y,\alpha})
+(\epsilon_\mathrm{p}-\mu)\hat X _{\mathbf{n},\alpha}]\nn\\&
+\hat Y _{\mathbf{n},\alpha}^\dagger [ -t_{sp}(-\hat S_{\mathbf{n},\alpha}+\hat S_{x,y+1,\alpha})\nn\\&
-t_{pp}(\hat X_{\mathbf{n},\alpha}-\hat X_{x-1,y,\alpha}
-\hat X_{x,y+1,\alpha}+\hat X_{x-1,y+1,\alpha})\nn\\&
-t_{pd}(\hat D_{\mathbf{n},\alpha}-\hat D_{x,y+1,\alpha})
+(\epsilon_\mathrm{p}-\mu)\hat Y _{\mathbf{n},\alpha}]\nn\\&
-J_{sd}\sum_\beta \left.
\hat S_{\mathbf{n},\beta}^\dagger \hat D_{\mathbf{n},\alpha}^\dagger
\hat S_{\mathbf{n},\alpha} \hat D_{\mathbf{n},\beta}\right\},
\label{RealSpaceHamiltonian}
\end{align}
where  $\hat S_{\mathbf{n},\alpha}$,
$\hat D_{\mathbf{n},\alpha}$,
$\hat X_{\mathbf{n},\alpha}$, and
$\hat Y_{\mathbf{n},\alpha}$ are Fermi annihilation operators 
at site or unit cell
$\mathbf{n}=(x,y)$, 
$x,\,y=0,\pm 1,\pm 2,\pm 3,\dots$ of the CuO$_2$ lattice,
$\mu$ is the chemical potential,
$\alpha$ and $\beta$ are spin indices.
and $J_{sd}$ is the anti-ferromagnetic Shubin-Zener-Kondo exchange amplitude.
For the operator of electron number analogously we have
\be
\hat N=-\partial_\mu \hat{H}^\prime. \nn
\ee
We would like to point out that the LCAO approach
is suitable tool to treat the Mott 
metal-insulator transition, see e.g. Ref.~\cite{Abrikosov:03} and references therein.

In momentum representation
\be
\hat\Psi_{\mathbf{n},\alpha}
\equiv
\begin{pmatrix}
\hat D_{\mathbf{n},\alpha}\\
\hat S_{\mathbf{n},\alpha}\\
\hat X_{\mathbf{n},\alpha}\\
\hat Y_{\mathbf{n},\alpha}
\end{pmatrix} \!
= \! \frac1{\sqrt{N}}\sum_{\mathrm{b},\mathbf{p}}
\mathrm{e}^{\mathrm{i}\mathbf{p}\cdot\mathrm{n}}
\begin{pmatrix}
D_{\mathrm{b},\mathbf{p}}\\
S_{\mathrm{b},\mathbf{p}}\\
\mathrm{e}^{\mathrm{i}\varphi_x}X_{\mathrm{b},\mathbf{p}}\\
\mathrm{e}^{\mathrm{i}\varphi_y}Y_{\mathrm{b},\mathbf{p}}
\end{pmatrix}
\hat{c}_{\mathrm{b},\mathbf{p},\alpha}
\label{n_p}
\ee
where the phases 
\be
\mathrm{e}^{\mathrm{i}\varphi_x}=\mathrm{e}^{\mathrm{i}p_x/2},\qquad
\mathrm{e}^{\mathrm{i}\varphi_y}=\mathrm{e}^{\mathrm{i}p_y/2} \nn
\ee
are chosen to provide real values for the eigenfunctions of
Hamiltonian \eqref{RealSpaceHamiltonian} in real space representation.
Further,
we will omit details of the standard substitution of the plane waves
\eqref{n_p}.
From the technical point of view, we obtain sums
over the different momenta with conserved total one
\be
\sum_{\substack{
\mathbf{p}^\prime,\,\mathbf{q}^\prime,\,\mathbf{p},\,\mathbf{q}\nn \\
\alpha, \, \beta}} \! \! \!
\delta_{\mathbf{p}^\prime+\mathbf{q}^\prime ,\,\mathbf{p}+\mathbf{q}}\,
S_{\mathbf{q}^\prime} D_{\mathbf{p}^\prime}
\langle
{\color{black}\hat c^\dagger_{\mathbf{q}^\prime\beta} \hat c^\dagger_{\mathbf{p}^\prime\alpha}}
{\color{black}\hat c_{\mathbf{p}\alpha}\hat c_{\mathbf{q}\beta}}
\rangle
S_\mathbf{p}D_\mathbf{q}.
\label{H_sd_1}
\ee
The BCS reduction of the Hamiltonian requires
to take into account only annihilation operators with opposite momenta
and simultaneously in self-consistent approximation 
to approximate the averaged product  of creation
and annihilation operators with the product of averaged 
two creation and two annihilation operators
\begin{align}
&
\delta_{\mathbf{p}^\prime+\mathbf{q}^\prime ,\,\mathbf{p}+\mathbf{q}}\,
S_{\mathbf{q}^\prime} D_{\mathbf{p}^\prime}
\langle
{\color{blue}\hat c^\dagger_{\mathbf{q}^\prime\beta} \hat c^\dagger_{\mathbf{p}^\prime\alpha}}
{\color{red}\hat c_{\mathbf{p}\alpha}\hat c_{\mathbf{q}\beta}}
\rangle
S_\mathbf{p}D_\mathbf{q}\nn\\
&
\approx
\delta_{\mathbf{q}^\prime+\mathbf{p}^\prime,0}\,
\delta_{\mathbf{q}+\mathbf{p},0} \, \chi_{\mathbf{p}^\prime} 
\langle{\color{blue}\hat c^\dagger_{-\mathbf{p}^\prime\overline\alpha} 
\hat c^\dagger_{\mathbf{p}^\prime\alpha}}\rangle    
\langle{\color{red}\hat c_{\mathbf{p}\alpha}\hat c_{-\mathbf{p}\overline\alpha}}\rangle
\chi_{\mathbf{p}} 
\nn
\end{align}
Additionally in these anomalous averages, we have to perform 
a Fermi liquid (FL) reduction
\begin{align}
&
\delta_{\mathbf{p}^\prime+\mathbf{q}^\prime ,\,\mathbf{p}+\mathbf{q}}\,
S_{\mathbf{q}^\prime} D_{\mathbf{p}^\prime}
\langle{\color{red}\hat c^\dagger_{\mathbf{q}^\prime\beta}} 
{\color{blue}\hat c^\dagger_{\mathbf{p}^\prime\alpha}\hat c_{\mathbf{p}\alpha}}
{\color{red}\hat c_{\mathbf{q}\beta}}\rangle
S_\mathbf{p}D_\mathbf{q}\nn\\
&
\approx
\delta_{\mathbf{q}^\prime,\mathbf{q}}\,
\delta_{\mathbf{p}^\prime,\mathbf{p}} \,     
\chi_{\mathbf{p}}
\langle{\color{blue}\hat c^\dagger_{\mathbf{p}^\prime\alpha}\hat c_{\mathbf{p}\alpha}}\rangle
\langle{\color{red}\hat c^\dagger_{\mathbf{q}^\prime\beta}} 
{\color{red}\hat c_{\mathbf{q}\beta}}\rangle\chi_{\mathbf{q}}\nn
\end{align}
where
\be
\chi_\mathbf{p}\equiv S_\mathbf{p}D_\mathbf{p},\qquad
S_\mathbf{p}=S_{-\mathbf{p}}.\nn
\ee
One of the main result we report here is the coincidence of BCS~\cite{MishonovPenev:11} 
and FL multiplicatively separable kernels of the reduced Hamiltonians. 
Moreover, two fermion operators in averaging brackets 
have to be considered as averaged number of particles.

\section{Notions and notations}\label{notions}
The critical temperature $T_c$ and the superconducting gap
$\Delta$ are calculated within the standard BCS approach
\begin{equation}
\label{BCS_gap_equation}
2J_{sd}\,\int_{0}^{2\pi}\int_{0}^{2\pi}
\frac{\chi_\mathbf{p}^2}{2E_\mathbf{p}}
\tanh\left(\frac{E_\mathbf{p}}{2T}\right)
\frac{\md p_x\md p_y}{(2\pi)^2}
=1,
\end{equation}
with
\begin{align}
	E_\mathbf{p}\equiv & \ \sqrt{\eta_\mathbf{p}^2+\Delta_\mathbf{p}^2},\nn\\
	\eta_\mathbf{p}\equiv & \epsilon_\mathbf{p}-\eF,
\qquad \Delta_\mathbf{p}=\Xi(T)\,\chi_\mathbf{p}, \notag
\end{align}
where
$\Xi(T)$ is the superconducting order parameter,
$\epsilon_\mathbf{p}$ is energy of the conduction band,
which obeys the secular equation
\be
\mathcal{A}xy+\mathcal{B}(x+y)+\mathcal{C}=0.
\label{spectrum}
\ee
Here, we have introduced
\be
x=\sin^2\left(\frac{p_x}{2}\right),  \qquad y=\sin^2\left(\frac{p_y}{2}\right) \nn
\ee
and 
\begin{align}
\mathcal{A}\left(\epsilon\right) & = 32 \tau_{sp}^2 \left(2 t_{pd}^2 + t_{pp} \varepsilon_d \right),
\quad \tau_{sp}^2=t_{sp}^2-\frac12\varepsilon_s t_{pp},
\nn\
\\
\mathcal{B}(\epsilon) & = -4\varepsilon_\mathrm{p}(t_{sp}^2\varepsilon_d+t_{pd}^2\varepsilon_s), \nn
\\
\mathcal{C}(\epsilon) & = \varepsilon_d\varepsilon_s\varepsilon_\mathrm{p}^2, 
\nn
\end{align}
with
$\varepsilon_s=\epsilon-\epsilon_s$,
$\varepsilon_d=\epsilon-\epsilon_d$, 
$\varepsilon_\mathrm{p}=\epsilon-\epsilon_\mathrm{p}$.
The main detail of the LCAO-$J_{sd}$ theory for the electron processes in CuO$_2$ plane 
is the function $\chi_\mathbf{p}$, the separable exchange interaction
\begin{align}
\chi_\mathbf{p} = & \; S_\mathbf{p}D_\mathbf{p}
=4\varepsilon_\mathrm{p}t_{sp}t_{pd}(x-y)  \nn \\
& \times \left[
\varepsilon_s\varepsilon_\mathrm{p}^2-4\varepsilon_\mathrm{p}t_{sp}^2\,(x+y)
+32t_{pp}\tau_{sp}^2\,xy
\right]\nn\\
& \times\left\{
\left[4\varepsilon_\mathrm{p}t_{sp}t_{pd}\,(x-y)\right]^2\right.\nn\\
&\qquad
+\left[\varepsilon_s\varepsilon_\mathrm{p}^2-4\varepsilon_\mathrm{p}t_{sp}^2\,(x+y)
+32t_{pp}\tau_{sp}^2\,xy \right]^2\nn\\
&\qquad
+4x\left[(\varepsilon_s\varepsilon_\mathrm{p}-8\tau_{sp}^2y)t_{pd}\right]^2\nn\\
&\qquad
\left.+4y\left[(\varepsilon_s\varepsilon_\mathrm{p}-8\tau_{sp}^2x)t_{pd}\right]^2
\right\}^{-1},
\end{align}
where $S_\mathbf{p}$ and $D_\mathbf{p}$ are the amplitudes for the
band electron to be in the Cu4$s$ and Cu3$d_{x^2-y^2}$ orbitals, respectively.
In other words, $\chi(p_x,p_y)$ is the magnitude of \textit{s-d} hybridization,
the main ingredient of the matrix elements of \textit{s-d} exchange interaction.
This hybridization amplitude $\chi_\mathbf{p}$ enters into the 
interaction kernel
\be
f(\mathbf{p},\mathbf{q})\equiv-2J_{sd}\chi_\mathbf{p}\chi_{\mathbf{q}}
\label{f(p,q) BCS}
\ee
which is one and the same in the BCS and Fermi liquid reductions
of the exchange \textit{s-d} Hamiltonian.
Let us note that this result is corroborated by the
proof of Pokrovsky~\cite{Pokr:61,PokrRiv:63} that in the weak coupling limit of
the BCS theory any arbitrary pairing can be approximated by a
separable kernel.
In this case $\chi_\mathbf{p}$ is the eigenfunction
of the pairing kernel corresponding to maximal in modulus 
eigenvalue. 
For more computational details the interested reader may consult an
unabridged version of the present study~\cite{arXiv_Pokrovsky}.
In the following we give a microscopic explanation of the hot spots observed 
by ARPES experiments and postulated phenomenologically in Refs. \cite{Hlubina:95,Ioffe:98}. 
In some sense this is a hint for the importance of the \textit{s-d} exchange interaction
in shaping the electronic properties of the CuO$_2$ plane. Now we are
in position to address analytically the gap anisotropy and the kinetics of the
normal phase that has been postulated in the past. We will present
some results of the self-consistent treatment of this $J_{sd}\,$-LCAO
Hamiltonian \eqref{RealSpaceHamiltonian} and compare to ARPES
data.

Before proceeding further with our analysis, we
would like to point out that the ``Tight binding'' and ``LCAO''
methods are to some extent equivalent. Generally, the tight binding
method is mainly used in a more mathematical physics context,
while LCAO
suggests that the parameters of the lattice Hamiltonian could be
evaluated starting from the atomic structure and the corresponding wave functions.
For example,
the transfer (or hopping integrals) $t_{pp}$, $t_{pd}$ and $t_{sp}$
can be evaluated as surface integrals of the wave functions of neighboring atoms.
The corresponding problem for H$_2^+$ ion is provided in many textbooks on quantum mechanics, for example see Ref.~\cite{LL3}.
Following the same reasoning, we obtain for the Hubbard $U$ integral 
\begin{align}
U_{dd}& = \left\langle\frac{e^2}{r_{12}}\right\rangle  \nn \\
& = \iint\frac{e^2}{r_{12}}\left\vert\psi_{\mathrm{Cu} 3d}\left(\mathbf{r}_1\right) \right\vert^2\left\vert\psi_{\mathrm{Cu} 3d}\left(\mathbf{r}_2\right) \right\vert^2\md^3\mathbf{r}_1\md^3\mathbf{r}_2\nn
\end{align}
describing the Coulomb repulsion of two electrons 
in one Cu atom \cite[Eq.~(19), page 81]{Mott:90}. 
As the Cu3$d$ orbital is the closest of all orbitals to the nucleus
from the 4 band LCAO model,
the corresponding $U_{dd}$ integral is the largest.
In brief, the lattice Hamiltonian accounts for the atomic structure via the atomic wave functions.
This trivial consideration above is readily applied to 
the Mott transition
or charge transfer Mott transition, as well as the role of the
exchange interaction are discussed very recently in Ref.~\cite{Barisic:22}.
The use of LCAO for Mott transition and related topics is known for at least half a century,
see e.g. the monograph by Mott~
\cite[Eq.~(10), p.~9; Eq.~(12), p.~12, 
        Eq.~(26), p.~25; Eq.~(19), p.~81, 
        Eq.~(24), p.~100, Eq.~(40), p.~116, p.~129]{Mott:90}
on metal-insulator transitions.
Last but not least, the Mott transition is not closely related to our present study.
Hot/cold spots and BCS approach are better applicable to over-doped cuprates,
which are more or less normal metals.

\section{Results}
\label{sec:Res}

The performed calculations in the framework of the
derived \textit{s-d} LCAO Hamiltonian
were performed with values of the single site energies
$\epsilon_d=0$~eV, $\epsilon_p=-0.9$~eV, $\epsilon_s=4.0$~eV
and the hopping integrals
$t_{sp}=2.0$~eV, $t_{pd}=1.5$~eV~\cite{Andersen:95,Pavarini:01},
$t_{pp}=0.2$~eV~\cite{Mishonov:96}.
The in-plane lattice constant $a_0=3.6$~\r{A}, while 
the filling factor $f_h = 0.58$ is chosen to
correspond to the optimally hole doped cuprates.
Before addressing our final purpose of hot/cold spots of layered cuprates,
we rederive some well-known results
for their superconducting properties. 
In order to explain the normal state phenomenology, we use 
the \textit{s-d} exchange model that is able to provide an adequate
explanation of the gap symmetry and the anisotropy.

\subsection{Evaluation of the critical temperature $T_c$}

The remarkable correlation between the critical temperature $T_c$ and the
electronic parameter $r$ \cite{Pavarini:01}
covers the whole temperature range of 
cuprate high-$T_c$ superconductivity and its explanation is an indispensable 
ingredient of the theory of high-$T_c$ superconductivity.
The dimensionless parameter is defined by \cite{Pavarini:01}
\begin{align}
r\equiv\frac1{2(1+s)},\quad 
s(\eF)\equiv
	\dfrac{(\epsilon_s-\eF)(\eF-\epsilon_\mathrm{p})}{(2t_{sp})^2}.
\label{reF}
\end{align}
Here $\epsilon_s$, $\epsilon_d$, and $\epsilon_\mathrm{p}$
are single site energies for Cu4$s$, Cu3$d_{x^2-y^2}$ and O2$p$
atomic levels, 
and $\eF$ is the Fermi energy. 
Following Ref.~\cite{Pavarini:01} 
the transfer integrals between these four atomic orbitals 
are denoted by $t_{pd}$, $t_{sp}$, $t_{pp}$.

\begin{figure}[ht]
\centering
\includegraphics[width=\columnwidth]{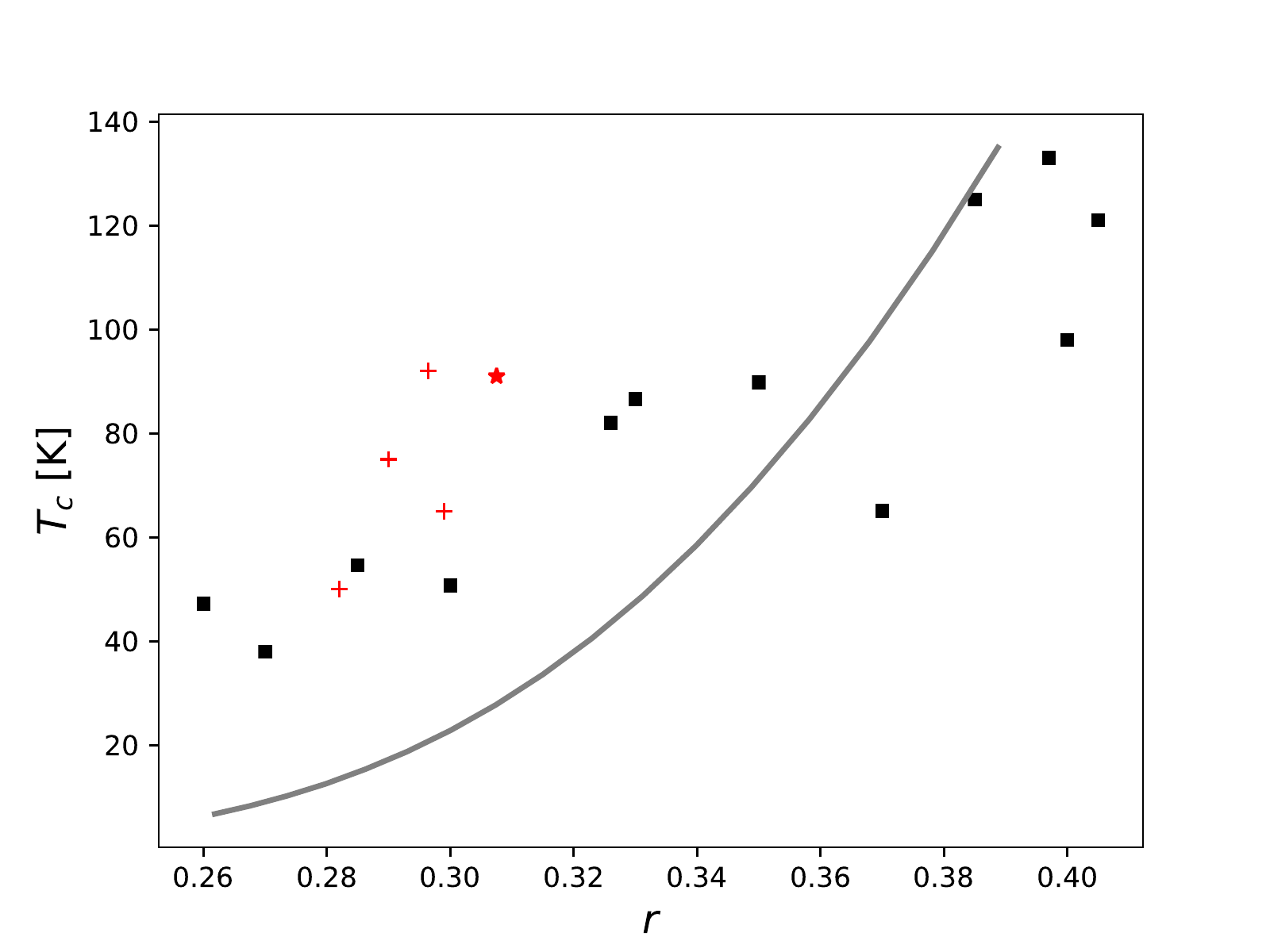}
\caption{Correlation between the $r$ parameter and $T_c$:
($\blacksquare$) Pavarini \textit{et al.}~\cite[Fig.~5]{Pavarini:01},
({\color{red}\textbf{+}}) Vishik \textit{et al.}~\cite{Vishik:10} 
for underdoped cuprates,
({\color{red}$\mathbf{\star}$}) 
Kaminski \textit{et al.}~\cite{Kaminski:05}
and Zonno \textit{et al.}~\cite{Zonno:21} from ARPES data;
the solid line~\cite{BJP:11} is our traditional BCS interpretation for \textit{s-d} exchange amplitude $J_{sd}$ approximately equal for all cuprates,
which is a fitting parameter of the
theory and can be used to determine the intensity of 
the scattering rate in the normal phase.
The value of $J_{sd}$ used to obtain the curve in the figure is 7.23~eV.}
\label{fig:rTc}
\end{figure}

The $r$-$T_c$ correlation for the Hamiltonian \eqref{RealSpaceHamiltonian} compared to
some experimental studies taken from Ref.~\cite{Pavarini:01} is depicted in Fig. \ref{fig:rTc}.
The continuous line in Fig.~\ref{fig:rTc} is the
result of our calculations according to the
gap equation \eqref{BCS_gap_equation} supposing that $J_{sd}$
is approximately the same for all layered cuprates at a fixed set of LCAO parameters.
%
Then for a fixed value of $J_{sd}$ varying only Cu4$s$ level $\epsilon_s$ we 
calculated $T_c$ from the same equation \eqref{BCS_gap_equation}.
For some novel materials the $\epsilon_s$ parameter is determined by
fitting the Fermi contour at fixed other parameters and then the
parameter $r$ is calculated according to \eqref{reF}.
This is an acceptable approximation
since $T_c$ and the shape
of the Fermi contour are very sensitive to $\epsilon_s$ as pointed out in
Ref. \cite{Pavarini:01}.

The band-structure correlations between the shape of the Fermi contour 
and the critical temperature for optimally hole doped cuprates reveal
two important conclusions:
1) we have a usual metal with single conduction band or
2) the electron conductivity is determined by a lower Hubbard band,
and the Fermi operator approximation provides a satisfactory accuracy
to be included in the standard BCS scheme.
The $d$-wave superconducting gap was also confirmed by ARPES measurements,
see Damascelli \textit{et. al.}~\cite[V.~Superconducting Gap, Figs.~46,~50]{Damascelli:03}
and references therein.
High-$T_c$ cuprates are doped Mott insulators,
but as it was pointed out by Lee, Nagaosa and Wen~\cite{Lee:06}:
beyond optimal doping (called the overdoped 
region), the normal metal properties gradually reappear.
Just in this region
the hot/cold spots phenomenology
\cite{Hlubina:95,Ioffe:98}
and the $r$--$T_c$ correlation are both applicable, which allows BCS treatment.
Recently Lee, Kivelson and Kim~\cite{Lee:16} used Bogolyubov-de Gennes 
BCS approach to analyse cold spots and glassy nematicity in underdoped cuprates.
This significantly extends the hot/cold spots phenomenology for cuprates;
it seems that the underdoped regime requires to take into account more
parameters in comparison to the
simple BCS picture of the overdoped regime.
That is why we have included the experimental data for underdoped 
cuprates in Fig.~\ref{fig:rTc} to the band trend
of the model under consideration.
%
The authors of Ref. \cite{Lee:16} show the existence of glassy charge order in the pseudogap phase of HTS cuprates.
Momentum and real space probes show charge density wave (CDW) order
with moderate finite correlation length.
Results from diffraction, local probes, and transport suggest nematic order. 
Most of the theory has focused
on long-range ordered states, or dynamically
fluctuating order parameter. Whereas glass order implies short-range
heterogeneities.
The authors concluded that the self-consistent BCS approach is an 
acceptable approximation for the study of
many characteristics of underdoped cuprates.

Electron band calculations of Fermi surfaces (contours) of the cuprates 
perfectly describe the experimentally observed ARPES Fermi contours,
while the realistic LCAO approximation of the \textit{ab initio} bands
are not that accurate,
since they neglect correlations.
Moreover in Hubbard model description of the electron structure 
the ratio of $t$-$t^\prime$ parameters is taken from
electron band calculations.
In any case the Cu4$s$ energy $\epsilon_s$ is included
in a unique way in the LCAO approximation of the band structure.

Let us recall the $J_{sd}$ exchange interaction is widely used to explore the
magnetism in d-metals, such as transition metals and their compounds.
In the present study, we have solved the corresponding integral
equations and have found that the $J_{pd}$ exchange amplitude gives a
gap anisotropy and hot/cold spots that do not agree with the
experimental data. 
It was shown \cite{Mishonov:94,Mishonov:98,Mishonov:02} that
the matrix elements of the $J_{pd}$ pairing added to the $J_{sd}$
pairing contribute only a de-pairing perturbation.
In this way the experimental data determine which exchange amplitude
is dominant.
In short, if we account for $J_{pd}$ or $J_{ps}$ in the BCS gap equation
the solution will possess again a crystal symmetry albeit a different one.

\subsection{From the Hubbard repulsion to the $s$-$d$ anti-ferromagnetic exchange interaction}

It is obvious that hybridization is a one body problem, yet to
determine the hybridization amplitude $\chi_\mathbf{p}$ it is
necessary to compute the matrix elements of the exchange interaction.
The \textit{s-d} exchange interaction was proposed eighty years ago
but up to date there are no reliable formulas for the calculation of
the exchange amplitude $J_\mathrm{sd}$  that is a
parameter of the theory.
The exchange amplitudes defining the physics of high-$T_c$
superconductors are also parameters of the theory.
To show that $J_\mathrm{sd}$ is anti-ferromagnetic in
nature, we start with
the well-known microscopic formula, see e.g. Ref.
\cite[Eq.~(7.17)]{WhiteGeballe:79} and the more recent textbook
\cite{White:07},
\begin{align}
	J = & 2\left\vert  V_{0\mathbf{k}_{_\mathrm
F}}\right\vert^2\frac{U}{E_0(E_0+U)} \notag \\
& \rightarrow - J_{sd}\simeq 2
\left\vert  t_{pd} \right\vert^2\frac{U_{dd}}
{(\epsilon_d -\eF)[(\epsilon_d -\eF)+U_{dd}]}.
\end{align}
Here we make some qualitative replacements:
1) the definition of the sign is a matter of convention,
we use positive $J_{sd}$ corresponding to singlet pairing and a tendency to anti-ferromagnetism;
2) the Hubbard $U$ is actually the Coulomb interaction $U_{dd}$ when two electrons are 
simultaneously on Cu3$d_{x^2-y^2}$ state;
3) the amplitude of electron transfer between Kondo impurity 
$V_{0\mathbf{k}_{_\mathrm F}}$ 
and an electron on the Fermi surface is just the lattice transfer integral $t_{pd}$;
4) the electron energy on the Kondo impurity $E_0$ is according to our
interpretation the energy level of the Cu3$d_{x^2-y^2}$ state $\epsilon_d$.
For more details, we recommend 
the monograph by White and Geballe~\cite[Chap.~7, Sec.~1]{WhiteGeballe:79}
and cited therein works by Anderson, Wolff, Schrieffer, and Wilson.
To proceed further, we use the Single Impurity Anderson Model (SIAM)~\cite{Anders:12}
for the virtual bound state in order to describe qualitatively 
the anti-ferromagnetism in the CuO$_2$ lattice,
where each Cu ion is viewed as Kondo impurity.
In short, this is not a proof but only 
a qualitative explanation why the phenomenology 
of the \textit{s-d} interaction can be successful for a simultaneous
description of the gap anisotropy and hot/cold phenomenology.

Nowadays it is known that the \textit{s-d} four-fermion interaction
explains fairly well the gap anisotropy in the overdoped cuprates
\cite{Anders:18,Anders:20,Anders:21a,Anders:21b}.
The study of the two-impurity Anderson model (TIAM) gives
new insights showing that the anti-ferromagnetic contribution to 
$J_\mathrm{RKKY}$ is determined by $\left(t^\mathrm{eff}\right)^2/U$ where $U$
denotes the Coulomb interaction~\cite{Anders:18} and RKKY (Ruderman-Kittel-Kasuya-Yosida).
The single-impurity problem was extended to lattice models~\cite{Anders:20},
multi-impurity Anderson models and periodic Anderson models~\cite{Anders:21a}
and multi impurity arrays~\cite{Anders:21b}. 
It would be interesting to explore this matter in CuO$_2$ to unveil
the influence of strong electron correlations on electron-electron scattering 
for overdoped cuprates.
In the next section we use the \textit{s-d} exchange interaction to describe
the anisotropy of the scattering rate in the normal phase of overdoped cuprates.

\subsection{Charge carriers scattering by density fluctuations.
Who could be blind to the beauty of the blue sky?
}

One of the main properties of the high-$T_c$ cuprates is their strong electrodynamic 
anisotropy, they possess conducting $a$-$b$ planes and almost dielectric behavior in the $c$-direction perpendicular to CuO$_2$ planes.
In the layered metal, the conducting CuO$_2$ layers (single or multiple)
are separated by insulating layers.
In some sense the $c$-direction perpendicular to the layers can be
dubbed dielectric direction.
Conducting layers (CuO$_2$)$_2$, 
double for YBa$_2$Cu$_3$O$_{7-\delta}$ 
and Bi$_2$Sr$_2$Ca$_1$Cu$_2$O$_8$,
 serve like plates of a plane capacitor.
In 1907 Albert Einstein~\cite{Einstein:07}
pointed out that in a plane capacitor with a short circuit between its
plates, thermodynamic fluctuations of the electric voltage between the
plates is proportional to the thermal fluctuations of the electric field
$E_z$ perpendicular to the plates
\begin{align}
\frac12 CU^2=\frac12T,\quad U=c_0E_z,\quad C= \frac{\varepsilon_0 b_0^2}{c_0},
\end{align}
where 
$b_0^2$ is the area of the plates and we qualitatively assume $b_0\sim a_0$
to be equal to the lattice constant,
$c_0$ is the distance between conducting planes,
and $\varepsilon_0=1/4\pi$ 
or in SI $\varepsilon_0=1/4\pi c^2 \, 10^{-7}$.
Having a plane capacitor system, state-of-the-art statistical consideration
requires the fluctuation of the electrical field $E_z$ to be taken into account.
For a layered metal with weak coupling between layers,
fluctuations of the transverse electric field have to be taken into account.
On the other hand the thermodynamic fluctuations of the electric field
are proportional
to the thermodynamic fluctuation of the two dimensional electron density
of metallic CuO$_2$ layers $q_e \, \delta n_\mathrm{2D}=\varepsilon_0E_z$.
The thermodynamic fluctuations of two dimensional electron density $\delta n_\mathrm{2D}$
are a corner stone of the theory of Ohmic resistivity $\varrho$.
We have strong pairing interaction in the superconducting phase,
but what happens in the normal phase?
A plane wave of charge carrier scatters off the density fluctuations 
by the exchange interaction.
This is analogous to the Rayleigh scattering~\cite{Rayleigh}
of the sunlight in the Earth atmosphere~\cite{Mishonov:00}.
Who could be blind to the blue sky~\cite{Mishonov:00}.
In short, 2D electrons scatter off by the fluctuation of the 2D electron density.
The linear dependence of the resistivity in the metallic $ab$-plane $\varrho_{ab}\propto T$ 
in this construction is just a demonstration of the
classical fluctuation of the electric field $E_z$ in the dielectric $c$-direction.
Statistics of waves scattered by thermal density fluctuations; 
we have common mechanism for the color of the blue
sky and Ohmic resistance of the layered transition metal perovskites.

Let us mark some details of this chain of considerations~\cite[Eqs.~(2.2) and (39.20)]{LL9}:
first we perform Fermi liquid reduction of the exchange Hamiltonian and
the reduced Fermi liquid Hamiltonian $ \hat H_{_\mathrm{FL}}$
determines the single particle spectrum
\begin{align}
\varepsilon(\mathbf{p},\mathbf{r})
= & \epsilon_\mathbf{p}
+\frac{\partial \hat H_{_\mathrm{FL}}}{\partial\hat
n_{\mathbf{p},\alpha}} 
 \rightarrow\  \epsilon_\mathbf{p}+\frac1{N}\sum_{\mathbf{q}.\beta}f(\mathbf{p},\mathbf{q}) n_{\mathbf{q},\beta}(\mathbf{r})\nn\\
= & \epsilon_\mathbf{p}
+\frac{(-2J_{sd})}{N}\chi_\mathbf{p}\sum_\mathbf{q}\chi_\mathbf{q} 
n_{\mathbf{q}}(\mathbf{r},t).
\label{Fermi_liquid_spectrum}
\end{align}
In the quasi-classical Wentzel-Kramers-Brillouin (WKB) approximation
in the spectrum \eqref{Fermi_liquid_spectrum}, we substitute
the thermal fluctuations of the Fourier components $\delta n_{\mathbf{q}}$
of the 2D electron density proportional to the whole electron density
at some space point $\mathbf{r}$ via
\be
\frac1{N}\sum_\mathbf{q}\chi_\mathbf{q} \, \delta n_{\mathbf{q}}(\mathbf{r},t)
\simeq \delta n(\mathbf{r}).
\ee
The thermodynamic fluctuations suggest that the dispersion
of the 2D electron density is proportional to the temperature 
\be
\langle [ \delta n(\mathbf{n}) ]^2 \rangle \propto T.
\label{propto T}
\ee
Thus,
using the space $\mathbf{r}$ dependent 
component of the Fermi liquid correction to the spectrum
\eqref{Fermi_liquid_spectrum} as a random scattering potential
with the aid of the second Fermi golden rule of the perturbation theory,
we obtain scattering rate proportional to the temperature and
square of the hybridization amplitude 
$\Gamma_\mathbf{p}\propto T\chi_\mathbf{p}^2$. 
As the \textit{s-d} hybridization function may be approximated with an acceptable
accuracy with a single sinusoidal, introducing
$\tilde\theta=\theta-\tfrac\pi2$
for the angular dependence of the scattering rate 
\begin{align}
-\mathrm{Im} (\epsilon_\mathbf{p})
\propto\Gamma_\mathbf{p}
&=\frac{\Gamma_0}4\sin^2(2\tilde\theta)+\frac1{\tau_0}
\approx \Gamma_0\tilde\theta^2+\frac1{\tau_0},
\nn
\end{align}
where $\Gamma_0=k_1T$,
we obtain the results of the Ioffe and Millis 
phenomenology~\cite[Eqs.~(4-5)]{Ioffe:98}.
Analyzing kinetics of the normal phase 
Ioffe and Millis~\cite{Ioffe:98} postulate a separable kernel~\cite[Eq.~(22)]{Ioffe:98}
which naturally can be derived from the \textit{s-d} interaction.

The idea of thermal fluctuations of the electric charge 
and the associated density fluctuations of the two dimensional charge carriers density
can be most easily interpreted in terms of a plane capacitor model.
Two layer unit cell cuprates have in this sense a plane capacitor performed 
by the double (CuO$_2$)$_2$ layer.
However, even around a single CuO$_2$ plane thermal fluctuations of the electrostatic
potential will be in some sense uniform. 
Every 2D fluctuating mode will have again $\frac12 T$ energy according to the equipartition theorem.
The only condition for the applicability of the classical statistics is
that the temperature should be higher than the typical
frequency of the eigenmodes,
but 2D plasmons are gapless. 
For thin films in the superconducting phase the plasmon frequency can be significantly smaller than the superconducting gap and temperature,
as it was theoretically predicted~\cite{MishonovGroshev:00}
and later experimentally confirmed~\cite{Buisson:94}.
The idea that linear Ohmic  thermal resistance 
can be created by thermal fluctuations of the electric potential
may be exploited in many physical phenomena.
Only the the corresponding electrostatic task will be slightly different.
We conclude that it is praiseworthy to perform the relevant calculations
to the specific system under consideration, here however, we continue with the simplest
implementation of a plane capacitor.
The comparison of the detailed electrostatic problem to experimentally
observed Ohmic resistance can be considered as a final explanation of this long standing problem.
Appropriate layered structures with single layer copper oxide 
planes are available for many years~\cite{Logvenov:09}.
In the opposite case of perovskites with moderate anisotropy
when $c$-polarized plasmons have frequency higher than the temperature
the electric field-density fluctuations are frozen and Ohmic resistivity
$\varrho_{ab}\propto T^2$ according to Baber~\cite{Baber:37},
and Landau-Pomeranchuk~\cite{Landau:36} theory.
For a qualitative consideration and state-of-the-art calculation 
see the monographs by Mott~\cite[page~72]{Mott:90} 
and Lifshitz and Pitaevskii~\cite[Sec.~1]{LL9} and \cite[Sec.~75, 76]{LL10}.

\subsection{Some remarks on ARPES}
Taking into account the thermodynamic fluctuations of $E_z$ 
according to the
Herapath-Waterston equipartition
theorem~\cite{Herapath:21,Waterston:51,Waterston:92},
we obtain that the resistivity is proportional to the
temperature, i.e. $\varrho\propto T$.
Formally, the scattering of the density fluctuations is described by
an imaginary correction to the electron spectrum
\begin{equation}
\epsilon_\mathbf{p}\rightarrow \epsilon_\mathbf{p} 
=\Sigma^\prime(\mathbf{p},\omega)
+\mathbf{i}\Sigma^{\prime\prime}(\mathbf{p},\omega),
\end{equation}
where
\begin{equation*}
-\Sigma^{\prime\prime}=\Gamma_\mathbf{p}
=k_1T\,\frac{\Gamma_0}4\sin^2(2\tilde\theta)+\frac1{\tau_0}.
\end{equation*}
Here we wish to recall the well-known relation between the Green function $G$
and the self-energy $\Sigma$, and the one particle spectral function $A$
\begin{align}
&\notag
G(\mathbf{p},\omega)=\frac1{\hbar\omega-\epsilon_\mathbf{p}-\Sigma}\,,\\
&
A(\mathbf{p},\omega)=
-\frac{1}{\pi}
\frac{\Sigma^{\prime\prime}}
{
[\hbar\omega-\epsilon_\mathbf{p}-\Sigma^\prime]^2
+[\Sigma^{\prime\prime}]^2}\,,\notag\\
&
\int_{-\infty}^\infty A(\mathbf{p},\omega) \mathrm{d}(\hbar\omega)=1,
\notag
\end{align}
see the well known monographs~\cite[Sec.~14 Self-energy function]{LL9},
\cite[Sec.~10 Dyson equation]{AbrGorDzya}
and the review~\cite[Eqs.~(13-20)]{Damascelli:03}.
The Boltzmann equation approach does not require
meticulously calculated spectral density $A(\mathbf{p},\omega)$,
but only a simple analytical approximation for the imaginary component of the self-energy
\be
\frac1{\tau_\mathbf{p}}=\Gamma_\mathbf{p}
= Z_\mathbf{p} 
\vert \Sigma^{\prime\prime} \vert, 
\qquad  Z_\mathbf{p} \le 1.
\ee
For weak scattering of gas particles on static inhomogeneities
the kinetic approach gives the Lorentzian  approximation
\be
A(\mathbf{p},\omega)=\frac1\pi\frac{\hbar/\tau_\mathbf{p}}
{(\hbar\omega-\varepsilon_\mathbf{p})^2
+(\hbar/\tau_\mathbf{p})^2}.
\ee
We recall the basic notion of electron spectra just to establish a bridge between the CuO$_2$ plane thermal fluctuations of the electric potential and ARPES spectra.
An exact extraction of the width of the approximating Lorentzian from experimental data is far beyond the purpose of this initial study.

In the nice review by Lee, Nagaosa and Wen~\cite{Lee:06}
on the physics of 
high-$T_c$ superconductors, such as doped Mott insulators, it is emphasized 
that the normal state of the optimally doped ones exhibits unusual properties.
Linear in $T$ resistivity is quoted as a nice
illustration of non-Fermi liquid behavior since the early days of
high-$T_c$ superconductivity.
We wish to comment that  this relationship is valid even for optimal $T_c$,
because this linearity is a simple proof of the applicability of the
equipartition theorem for transverse electric field in layered metals.
Nothing \textit{strange} that statistical physics is applicable to a layered \textit{metal};
this situation is known as ``strange metal'', i.e. applicability of the Fermi liquid theory
to layered metals.
A non-Fermi liquid is just a layered Fermi gas with exchange
interaction and electric field between  the layers.
This problem however is far from the linear resistivity
solution
explained by Rosch~\cite{Rosch:00} as a property of nearly antiferromagnetic metals 
close to the quantum critical point.
Later on Lee~\cite{Lee:21} showed that the low temperature $T$-linear 
resistivity may be traced back to umklapp scattering from a critical mode.

\begin{figure}[h!]
\centering
\includegraphics[width=\columnwidth]{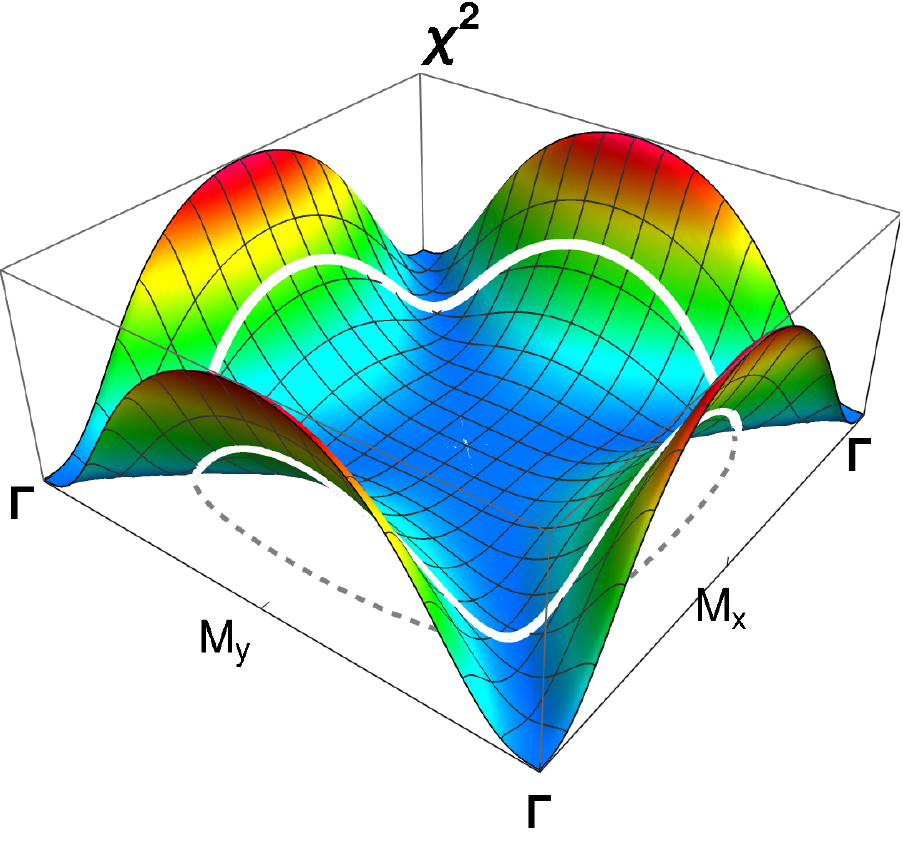}
\caption{The hybridization probability $\chi_\mathbf{p}^2=S_\mathbf{p}^2D_\mathbf{p}^2$.
Heights around the $M$ points correspond to hot spots,
while the navigation channels in the deep blue sea along the diagonals between the $\Gamma$ points correspond to cold spots in agreement with both
Refs.~\cite{Hlubina:95,Ioffe:98}.
Fermi contour (dotted line) is projected on the  $\chi_\mathbf{p}^2$ surface.}
\label{fig:chi2}
\end{figure}

\begin{figure}[h!]
\centering
\includegraphics[width=0.7\columnwidth]{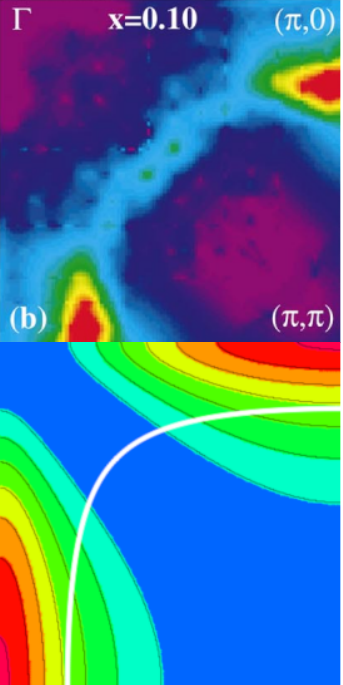}
\caption{Comparison of the scattering rate calculation in the framework of the 
\textit{s-d} exchange calculation (bottom)
with the ARPES data from
Armitage \textit{et al.}~\cite[Fig.~3~(b)]{Armitage:02} (top); reproduced with permission.
The momentum $\mathbf{p}$ dependence of the ARPES intensity
for electron energies is close to the Fermi one.
Continuous line in the theoretical calculation (bottom) denotes the Fermi contour.
The maxima in both figures correspond to hot spots from the phenomenology 
of Ref. \cite{Hlubina:95}.
The coincidence of hot and cold spots is the indispensable 
qualitative agreement before a detailed theory can be developed.}
\label{fig:spots}
\end{figure}

Concerning the applicability of the BCS theory to the high-$T_c$
superconductivity,
we wish to stress that recently 
Lee, Kivelson and Kim~\cite{Lee:16} have used
the de~Gennes--Bogolyubov approach to explain cold-spots and glassy 
nematicity in underdoped cuprates.
Comparing ARPES, scanning tunneling microscopy (STM)
and optical measurements with their BCS calculations, they observe 
consonance between cold-spot of glassy nematics and the gap nodes
of \textit{d}-wave superconductivity.
In the directions where the exchange interaction is zero, one may observe
the small contribution of the Coulomb scattering by density fluctuations which is also
$\propto T$.
In such a way, following a chain of standard approximations and
supposing that $\tau_\mathrm{cold}=\tau_0$, we arrive at the 
widely accepted anisotropy of the lifetime 
\begin{equation}
\frac1{\tau(\theta)}=\frac1{\tau_\mathrm{hot}}\cos^2(2\theta)+\frac1{\tau_\mathrm{cold}}.
\end{equation}
By introducing the averaged over the Fermi contour relaxation time
\begin{align}
	\tau_\mathrm{Drude}\equiv\ &\langle \tau(\theta)\rangle
=\int_0^{2\pi}\tau(\theta)\,\frac{\md \theta}{2\pi} \notag\\
	=&\dfrac{1}{\sqrt{\dfrac{1}{\tau_\mathrm{cold}\tau_\mathrm{hot}}
+\dfrac{1}{\tau_\mathrm{cold}^2}}}
\approx \sqrt{\tau_\mathrm{cold}\tau_\mathrm{hot}}\gg\tau_\mathrm{hot},\nn
\end{align}
the conductivity takes the standard Drude form
\begin{equation*}
\sigma_{ab}=
q_e^2n_e\frac{\tau_\mathrm{Drude}}{m_c},\qquad
\frac1{\tau_\mathrm{hot}}\equiv \frac{\Gamma_0}4,
\end{equation*}
where $m_c$ is the optical mass in CuO$_2$ plane;
$\tau_\mathrm{cold}$ is created by the Coulomb scattering and 
the small $\tau_\mathrm{hot}$ by the exchange one.
In order to check whether we are on the correct track 
we draw the hybridization 
probability $\chi_\mathrm{p}^2$ from \Fref{fig:chi2} together with 
ARPES data for the width of the 
spectral lines in \Fref{fig:spots}.
The comparison depicted in \Fref{fig:spots} shows a qualitative similarity, which is encouraging.
Originally the hot/cold spot phenomenology was proposed for optimally doped and hole overdoped cuprates. 
This phenomenology was supported by ARPES studies.
However, even for electron doped cuprates, see \Fref{fig:spots},
the lifetime anisotropy $1/\tau_\mathbf{p}$ is  qualitatively
similar,
which is a hint that the microscopic origin is the same.

\section{Discussion and conclusion}

We investigated the electronic properties of a generic
CuO$_2$ plane in the framework of Shubin-Kondo-Zener \textit{s-d} exchange interaction
that simultaneously describes the correlation between \textit{T}$\!_c$
and the Cu4\textit{s} energy. To achieve our goal,
we employed the Pokrovsky theory for anisotropic gap BCS superconductors.
We used a microscopic model to computed a multiplicatively separable kernel
able to simultaneously describe the scattering rate and the superconducting gap anisotropy.
Our theoretical approach reproduces the
phenomenological analysis of Refs. \cite{Hlubina:95,Ioffe:98}
performed to describe Angle Resolved Photoemission
Spectroscopy data.

We conclude that the electric charge fluctuations should be analyzed
in the framework of the standard theory of electromagnetic
fluctuations in continuous media~\cite[Chap.~8]{LL9} and
\cite[Chap.~6]{AbrGorDzya}.

The complete theory of ARPES is far beyond our reach,
we only wish to point out that hot/cold phenomenology can be
derived from a microscopic Hamiltonian describing the superconducting spectrum 
of the optimally doped and overdoped cuprates.
Moreover, it may seem that the named theory is also applicable to underdoped cuprates
if additional features like glassy nematicity is included
as it has been performed in Ref. \cite[Lee, Kivelson and Kim]{Lee:16}.

Roughly speaking, electrons have only an electric charge
and all exchange processes are the result of some projection or Hamiltonian reduction
taking into account the Fermi statistics.
Every exchange amplitude
has to be analyzed
in terms of its importance to a variety of viable physical processes.
For example, recently the Cu-Cu \textit{d-d} superexchange was used 
by Peng \textit{et al.}
for the interpretation of the observation 
of robust anti-nodal paramagnon modes following 
spin-wave-like dispersion by resonant inelastic x-ray scattering 
deep in the normal phase in overdoped cuprates~\cite{Peng:18}.
Moreover, zero sound modes were predicted~\cite{sound2022} for 
layered perovskites with ferromagnetic \textit{s-d}
exchange interaction.
Going back to the anti-ferromagnetic \textit{s-d} exchange interaction,
which dominates in many magnetic materials,
in the present work we suggest that it can be simultaneously
responsible for
the gap anisotropy of the superconducting phase and hot/cold spot phenomenology 
in the normal phase.
It would be worthwhile to derive the phenomenological 
\textit{s-d}~Hamiltonian starting from Hubbard model applied
to the CuO$_2$ plane.

\section*{Acknowledgments}
The authors are thankful to Patrick Lee for pointing out significant works on 
the $T$-linearity of conductivity. 
The stimulating correspondence with Katerina Piperova
is also highly appreciated.
Considerations with Mihail Mishonov and Evgeni Penev
at the early stages of this study are highly appreciated.

This work is partially supported by grant No K$\Pi$-06-H58/1
of the Bulgarian National Science Fund.
and Cost Action CA16218 --
Nanoscale coherent hybrid devices for superconducting quantum technologies.

\bibliography{../Article/Pokrovsky}

\end{document}